\documentclass[%
preprint,
superscriptaddress,					
 amsmath,amssymb,
 aps,
]{revtex4-1}

\usepackage{graphicx}						
\usepackage{dcolumn}						
\usepackage{bm}								
\usepackage{color}
\usepackage{comment}
\usepackage{environ}

\def\beq{\begin{equation}}
\def\eeq{\end{equation}}
\NewEnviron{eq}{%
\begin{equation}\begin{split}
\BODY
\end{split}\end{equation}
}
\def\bit{\begin{itemize}}
\def\eit{\end{itemize}}
\def\nn{\nonumber}

\def\te{\text}

\newcommand{\fig}[1]{Fig.~\ref{#1}}
\newcommand{\Fig}[1]{Figure~\ref{#1}}

\newcommand{\DR}{\textit{DRACO}} 
\newcommand{\LI}{\textit{LILAC}} 

\begin{document}
\title{Implementing a microphysics model in hydrodynamic simulations to study the initial plasma formation in dielectric ablator materials for direct-drive implosions}

\author{Arnab Kar}
\affiliation{Laboratory for Laser Energetics\\University of Rochester, Rochester, New York 14623, USA}
\author{S.~X. Hu}
\affiliation{Laboratory for Laser Energetics\\University of Rochester, Rochester, New York 14623, USA}
\author{G. Duchateau}
\affiliation{Universit\'{e} de Bordeaux-CNRS-CEA, Centre Lasers Intenses et Applications, UMR 5107, 351 Cours de la Lib\'{e}ration, 33405 Talence Cedex, France}
\author{J. Carroll-Nellenback}
\affiliation{Laboratory for Laser Energetics\\University of Rochester, Rochester, New York 14623, USA}
\author{P.~B. Radha}
\affiliation{Laboratory for Laser Energetics\\University of Rochester, Rochester, New York 14623, USA}

\begin{abstract}
A microphysics model to describe the photoionization and impact ionization processes in dielectric ablator materials like plastic has been implemented into the 1-D hydrodynamic code \LI\,for planar and spherical targets. 
At present, the initial plasma formation during the early stages of a laser drive are modeled in an \textit{ad hoc} manner, until the formation of a critical surface. 
Implementation of the physics-based models predict higher values of electron temperature and pressure than the \textit{ad hoc} model.
Moreover, the numerical predictions are consistent with previous experimental observations of the shinethrough mechanism in plastic ablators.
For planar targets, a decompression of the rear end of the target was observed, that is similar to recent experiments.
An application of this model is to understand the laser-imprint mechanism that is caused by nonuniform laser irradiation due to single beam speckle.
\begin{flushleft}
A. Kar, S.~X. Hu, G. Duchateau, J. Carroll-Nellenback, and P.~B. Radha, Phys. Rev. E \textbf{101}, 063202 (2020)
\end{flushleft}
\end{abstract}
\maketitle

\section{Introduction}

In direct-drive inertial confinement fusion (ICF), a spherical target is irradiated by laser beams to create the necessary conditions for fusion reactions~\cite{Craxton_Review}.
These targets contain the fusion fuel [deuterium (D) and tritium (T)] inside a plastic (CH) shell.
The laser energy on the target causes the plastic to ablate outward like the exhaust of a rocket.	
This ablation creates a reaction force on the remaining part of the capsule.
Ideally, one hopes that this should create a spherically symmetric implosion.
However, single beam speckle (laser imprint) can introduce perturbations that can compromise performance~\cite{Ishizaki}. 
In fact, the laser energy deposited from a single beam speckle penetrates through the target non-uniformly. 
This non-uniformity is intensified by the filamentation of the laser energy that leaves a damage track due to the self-focused laser radiation~\cite{Balmer}.
These variations act as seeds to Rayleigh--Taylor instability~\cite{Bodner,Hu_imprint} which grows exponentially. 
To create a uniform symmetric implosion for ignition, understanding and mitigating this laser imprint process is important~\cite{Draco,Radha_imprint,Peebles}.
As the laser beams irradiate the target, the target is ionized and a plasma is created around it. 
This coronal plasma determines the laser energy deposition on the target until a critical surface is established, and the target becomes opaque to the laser.
Here, corona refers to the region outside the highest mass density or electron density of the target before the critical surface is formed.
Once the critical surface is established, the region outside it is referred to as the corona.
After the critical surface formation, the subcritical underdense plasma absorbs the laser energy and transfers this energy through the electrons inside the critical surface to the ablation region.

Recently, it has been shown that the initial solid state of the target with specific electronic and optical properties has a notable impact on the subsequent plasma dynamics~\cite{Stuart_PRB,Duchateau_JAppPhys,Gamaly_PoP}.
It is important to implement a detailed and physically accurate model to understand the solid to plasma transition.
Thus, a microphysics model describing the response of the ablator material to the laser irradiation process on the target has been developed~\cite{Duchateau_Kar}.
The microphysics model incorporates a photoionization and impact ionization scheme that describes the transition of the solid ablator into plasma due to laser irradiation.

However, there are some limitations of the model as presented in Ref.~\cite{Duchateau_Kar} that are addressed in this paper. 
The model in that paper was a standalone model that focused on the target surface and did not include the dynamics of the ablation region or the interior of the target. 
Moreover, the model was not coupled with hydrodynamic equations that could study the plasma flow simultaneously with the laser energy deposition from a multibeam geometry, which is necessary to simulate an ICF target.
Developing an understanding of the plasma formation from the multibeam irradiation process on the target is crucial to address the laser-imprint mechanism. 
Implementing the microphysics model in a hydrodynamic code is the first step towards understanding the different physical processes that occur concurrently during the initial plasma formation. 
For example, it allows an estimation of the laser energy absorption over time while other relevant physical phenomena are captured in the hydrodynamic code. 
The impact ionization model in this paper was changed from the multiple rate equation (MRE) approach of Ref.~\cite{Duchateau_Kar} to the Drude model such that unknown material dependent parameter, like the rate of collisions between free and valence electrons and the rate for one-photon absorption in the conduction band, for CH are avoided. 
Also, the Drude model approach has the advantage of solving one differential equation unlike coupled differential equations in the case of MRE, which allows faster numerical calculations.

Traditionally, hydrodynamic codes have ignored this detailed transition mechanism from the solid-to-plasma state for the target. 
The hydrodynamic codes either assume that the material is ionized to start with and a critical electron density exists initially, or they adopt the ``cold-start" method where the laser energy is deposited on the surface of the target to generate a critical surface in an \textit{ad hoc} manner.
Both these strategies are incorrect from a physics perspective. 
Since radiation-hydrodynamic simulations form an essential component of our understanding of direct-drive ICF process, it is important to incorporate the microphysics model into the hydrodynamic codes to model the seeds of Rayleigh--Taylor growth including the initial solid state of the target.

In this paper, a more comprehensive version of the microphysics model~\cite{Duchateau_Kar} has been implemented into the 1-D hydrodynamic code \LI~\cite{Delettrez}.
We demonstrate the implications of the microphysics model in ICF through hydrodynamic simulations for both spherical and planar targets.
Unlike the \textit{ad hoc} model, the microphysics model shows laser energy absorption inside the target over time.
Additionally, the energy absorption causes the electron temperature inside the target to rise; subsequently, the pressure inside the target increases.
This is consistent with previous observations that the laser beam penetrates through the plastic and deposits energy inside the target, since plastic on the outermost layer of the target is transparent to UV laser light of 351--nm wavelength~\cite{Edgell}.
This phenomenon had not been captured in previous hydrodynamic simulations since the incident laser intensity was restricted to the target surface by creating a critical surface in an \textit{ad hoc} fashion.

This paper is organized as follows: In Sec.~\ref{model}, the microphysics model is described in detail along with the joule heating process, which determines the laser--energy absorption.
In Sec.~\ref{results}, the results from simulating a solid plastic sphere and a planar plastic foil are discussed, and the differences that arise between the \textit{ad hoc} model and the microphysics model are outlined.
Finally, in Sec.~\ref{summary}, the important results are summarized and future directions are outlined.

\section{Model}\label{model}

The details of the microphysics model to simulate the initial plasma formation are discussed in this section. 
This work focuses mainly on plastic or polystyrene ablators since they are commonly used ablator materials for direct-drive ICF targets. 
While the main features of the microphysics model developed in this paper should be applicable to other dielectric materials, although the material bandgap for the photoionization process is specific to CH in this paper.
The model does not require a specific range of parameters to be applicable besides the fact that it is focused on understanding the plasma dynamics before the critical surface formation.
Plastic (CH) is a dielectric material with a band-gap of 4.05 eV.
This makes solid plastic transparent to the UV laser light of 351--nm wavelength (or 3.53 eV), the wavelength of TW facilities like OMEGA~\cite{Boehly_OMEGA}.
Therefore, the laser energy shines through the target in the early stage of laser irradiation~\cite{Edgell}.
The solid ablator gradually transitions into a plasma state based on the laser energy absorption controlled by the microphysics model.

At present, hydrodynamic codes ignore the transparency of plastic to UV light, which is incorrect.
To establish a critical surface, the laser energy is deposited arbitrarily on the surface of the target only until a critical surface is established.
Another approach assumes that the plastic is partially ionized and the free electron density in the plastic is higher than the critical density to begin with.
This assumption is also incorrect since there are no free electrons in the conduction band of a dielectric material like plastic (no more than 10$^{10}$ cm$^{-3}$ which is negligible). 

\begin{figure}[htp]
\centering
\includegraphics[width=0.7\linewidth]{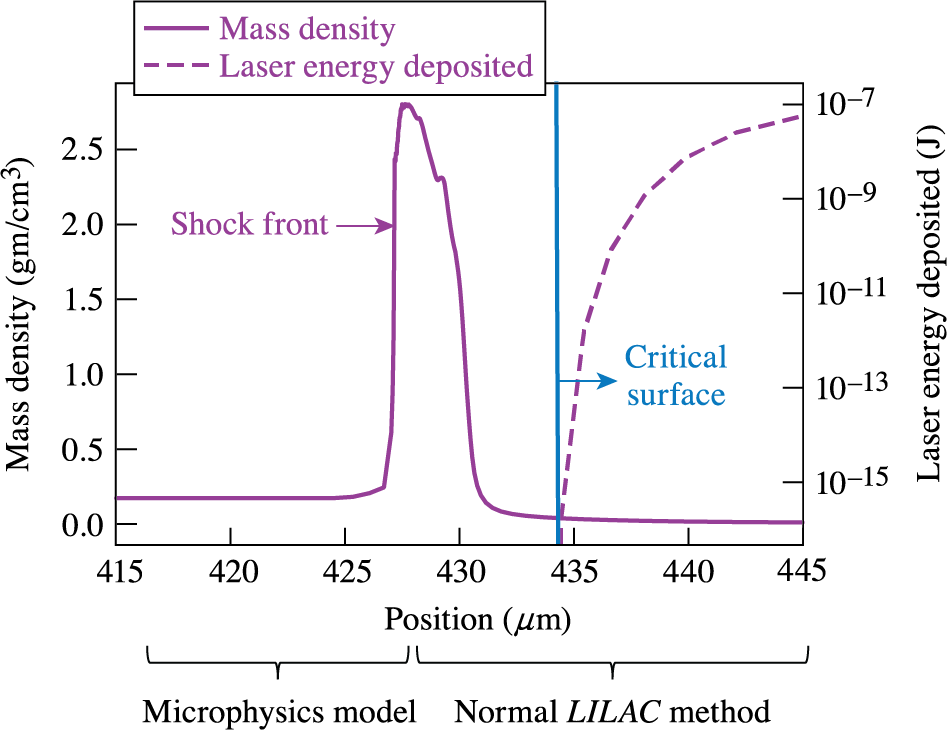}
\caption{An outline of the regions where the microphysics model and the normal \LI\,method is implemented after the critical surface is formed. In the figure, laser light is incident on the target from the right.}
\label{modelpicture}
\end{figure}

To overcome these inaccuracies and develop a physics-based model, a rate equation governing the free electron density of the electrons in the conduction band has been derived recently~\cite{Duchateau_Kar}.
This rate equation is coupled with a laser energy deposition scheme.
Based on the laser energy deposition, the plasma profile and the various physical quantities are determined.
This model governs the dynamics of the initial plasma formation from the solid throughout the target during the early stage of the irradiation, until a critical surface is created.
During this stage, the laser energy deposition is mediated by the joule heating mechanism of the electrons in the corona.
Once the critical surface forms, the material is ionized and assumed to be in the plasma state. 
After this time, the microphysics model dominates the plasma profile ahead of the shock front, while the normal \LI\,method for inverse bremsstrahlung absorption dominates the physics behind the shock front as demonstrated in \fig{modelpicture}.
There is no laser energy deposition represented by the dotted curve inside the critical surface.
The curly brackets below the figure show the different strategies used to model those regions.
The region behind the shock front (or the highest mass density front in the target) to the ablation region is modeled by the normal \textit{LILAC} model while the region ahead of the shock front is modeled using the microphysics model.

\subsection{Rate equation}

\begin{figure}[htp]
\centering
\includegraphics[width=0.7\linewidth]{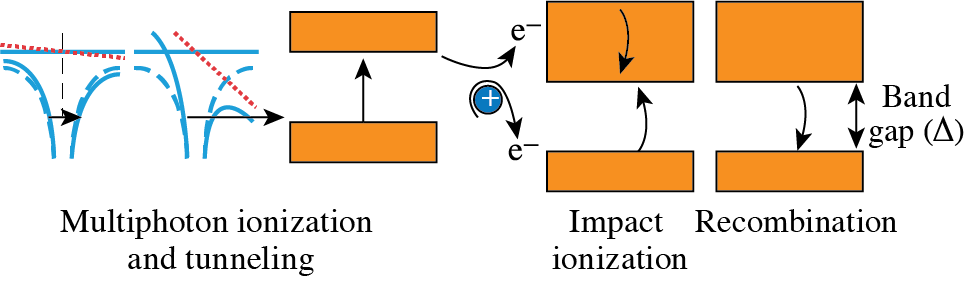}\quad
\caption{A schematic of the photoionization, impact ionization, and recombination processes that determine the free electron density in the conduction and valence band of the material.}
\label{schematic}
\end{figure}

For a dielectric material like CH, there are no free electrons in the conduction band at room-temperature conditions. 
\Fig{schematic} shows a schematic of three processes that excite electrons from the valence band to the conduction band of the material upon ionization. 
As the laser irradiates the target, the photon energy is absorbed by the electrons, which excites them from the valence band of the material to the conduction band.
This photoionization process dominates the early stage of the interaction.
The electrons can be excited by multiphoton absorption or tunneling depending on the strength of the laser field and the material parameters~\cite{Keldysh,Balling}.
These excited electrons can further absorb photon energy and be excited to higher energy states.
When they gain energies of the order of band-gap energy, they can collide with other valence electrons.
This further excites more valence electrons to the conduction band through impact ionization~\cite{Kennedy,DeMichelis}.
When such process takes place several times, the electron avalanche leads to an exponential increase in the electron density.
In this paper, the Drude model of impact ionization is implemented instead of the multiple rate equation approach used in Ref.~\cite{Duchateau_Kar}.
Electrons can also lose energy through the three-body or radiative recombination and recombine with the valence band~\cite{Zeldovich,Hinnov}.
These three processes are captured by the following rate equation, which determines the evolution of the free electron density $(n_{\te{fe}})$ in the conduction band:
\begin{eq}\label{rate}
\frac{\te{d}n_{\te{fe}}}{\te{d}t}=\left(1-\frac{n_{\te{fe}}}{n_{\te{vb}}}\right)W_{\te{PI}}+\eta\,n_{\te{fe}}-\frac{n_{\te{fe}}}{\tau_{\te{r}}}.
\end{eq} 
Here, $n_{\te{vb}}$ is valence-band electron density, $W_{\te{PI}}$ 	is photoionization rate, $\eta$ is impact ionization rate, $\tau_{\te{r}}$ is recombination time.
The rate equation captures photoionization and impact ionization that raises the electron density, while the recombination process leads to a loss of the electron density.
Processes such as diffusion that can also contribute to the free electron density change, are neglected here~\cite{Kroll}.
The coefficient in front of $W_{\te{PI}}$ ensures that the number of free electrons in the conduction band does not exceed the total number of valence electrons.
Based on the laser intensity, the multiphoton ionization process is dominant in the femtosecond time domain, while the impact ionization plays a significant role in the picosecond time scale when sufficient free electrons having energy larger than the band-gap energy have been generated to trigger the impact ionization~\cite{Rethfeld}.

\subsubsection{Photoionization}

The photoionization rate ($W_{\te{PI}}$) is determined using the Keldysh formula~\cite{Keldysh} based on the time-dependent electric field ($E_\te{L}$) of the laser (see details in Appendix~\ref{appen}).
The time-dependent laser pulse fixes $E_\te{L}$ for this photoionization process based on the incident laser intensity.
To implement the Keldysh formula, the ionization potential is assumed to be the same as the band-gap which is $\Delta=4.05$ eV for polymeric compounds like CH.
Based on the varying field strength, the photoionization rate $W_{\te{PI}}$ is determined in a time dependent manner.

\subsubsection{Impact ionization}

The Drude model provides a simple description of the motion of an electron in the presence of an electromagnetic field. 
In this model, the electron gains energy from the field upon collisions with other ions.
The energy absorption rate is~\cite{DeMichelis}:
\beq
\frac{\te{d}\epsilon}{\te{d}t}=\frac{e^2 E_\te{L}^2}{m_\te{e} \omega^2}\nu\frac{\omega^2}{(\omega^2+\nu^2)}.
\eeq
Here, $\omega$ is frequency of laser light, $E_{\te{L}}$ is electric field strength of the laser, $m_{\te{e}}$ and $e$	are electron mass and charge, $\nu$ is collisional frequency of electrons.
If the energy gained by the electron is approximately the band-gap energy of the material, they can then transfer that energy to another valence electron resulting in its ionization.
The rate of this impact ionization would be inversely proportional to the time taken to gain the energy equivalent to the band-gap of the material~\cite{Kennedy}.
\beq
\eta=\frac{e^2 E_\te{L}^2}{m_\te{e}\omega^2}\tau\frac{\omega^2}{(\omega^2\tau^2+1)\Delta},
\eeq
where $\tau =1/\nu$ is the electron collision time.

To evaluate the impact ionization rate, the collision frequency of the electrons must be determined.
The collision frequency $\nu$ of the electrons changes depending on the state of the matter.
Initially, when the material is cold, the electrons collide with the phonons of the solid state. 
In this phase, the collision frequency increases with the rise in ion temperature~\cite{Gamaly_1,Gamaly_2}.
Beyond 0.1 eV, roughly equivalent to the temperature at which the solid CH melts, the collisional frequency assumes a transitional value before the matter is completely ionized.
At this stage, $\nu$ is determined from the average of the Spitzer formula for the electron--ion collision~\cite{Atzeni} and the collisional frequency from the mean free path of the electrons.
The mean-free-path collisional frequency ($\nu_{\text{mfp}}$) of the electrons is determined from the thermal velocity of the electrons ($v_{\te{e}}$) and the ion-density ($n_{\te{i}}$). 
Eventually, as the material is ionized, the Spitzer collisional frequency is assigned for ion-temperatures ($T_{\te{i}}$) of more than 10 eV.
To summarize, the collisional frequency based on the ion temperature is given by
\begin{eq}
\nu= 
\begin{cases}
    \frac{T_{\te{i}}}{T_{\te{room temp}}}\times 10^{14}, & T_{\te{i}} < 0.1 \text{ eV},\\
    \frac{\nu_{\text{mfp}}\nu_{\text{Spitzer}}}{\nu_{\text{mfp}}+\nu_{\text{Spitzer}}}, & 0.1 \text{ eV} \leq T_{\te{i}} \leq 10 \text{ eV} \\
    \nu_{\text{Spitzer}}, & T_{\te{i}} > 10\text{ eV},
\end{cases},
\end{eq}
where 
\begin{eq}
\nu_{\text{mfp}}&=v_{\te{e}} n_{\te{i}}^{1/3}\\
\nu_{\text{Spitzer}}&=\frac{4}{3}(2\pi)^{1/2}\frac{\langle Z\rangle e^4m_en_{\te{fe}}}{(m_\te{e}k_\te{B}T_\te{e})^{3/2}}\log\Lambda\\
&=9.19\times 10^{-11}\frac{\langle Z\rangle n_{\te{fe}}}{T_{\te{e (keV)}}^{3/2}}\log\Lambda\quad \text{[CGS]}
\end{eq}
Here $\Lambda$ is determined from the maximum and minimum of the collision parameter~\cite{Eidmann}.
\begin{eq}
\Lambda&=\left[1+\left(\frac{b_{\te{max}}}{b_{\te{min}}}\right)^2\right]^{1/2},\\
b_{\te{min}}&=\max \left[\frac{\langle Z\rangle e^2}{k_{\te{B}} T_{\te{e}}},\frac{\hbar}{(m_{\te{e}} k_{\te{B}} T_{\te{e}})^{1/2}}\right],\\
b_{\te{max}}&=\left(\frac{k_{\te{B}} T_{\te{e}}}{m_{\te{e}}}\right)^{1/2}/\max(\omega,\omega_{\te{p}}),
\end{eq}
where $\omega_{\te{p}}$ is plasma frequency, $\langle Z\rangle$ is average atomic number in plasma, $k_{\te{B}}$ is Boltzmann constant and $T_{\te{e}}$ is electron temperature.
Note that the temperature regimes at which the electron collision frequency changes may change based on the material under consideration.

\subsubsection{Recombination}

The third term in the rate equation is the recombination term. 
The three-body recombination time is given by~\cite{Dawson,Zeldovich}:
\beq
\tau_{\te{r}}=1.8\times10^{26} \frac{T_{\te{e(eV)}}^{9/2}}{n_{\te{fe}}^2}
\eeq
for ion temperatures greater than 0.1 eV. 
For ion temperatures below that, the recombination time is set as 1 ps in the solid phase~\cite{Duchateau_recomb,Gattass}.
The three-body recombination has the opposite effect of the impact ionization process since some of the excited electrons in the conduction band fall back into the valence band due to mutual collisions~\cite{Apostolova}.

\subsection{Energy absorption mechanism}

\begin{figure*}[!htpb]
\centering
\includegraphics[width=0.9\linewidth]{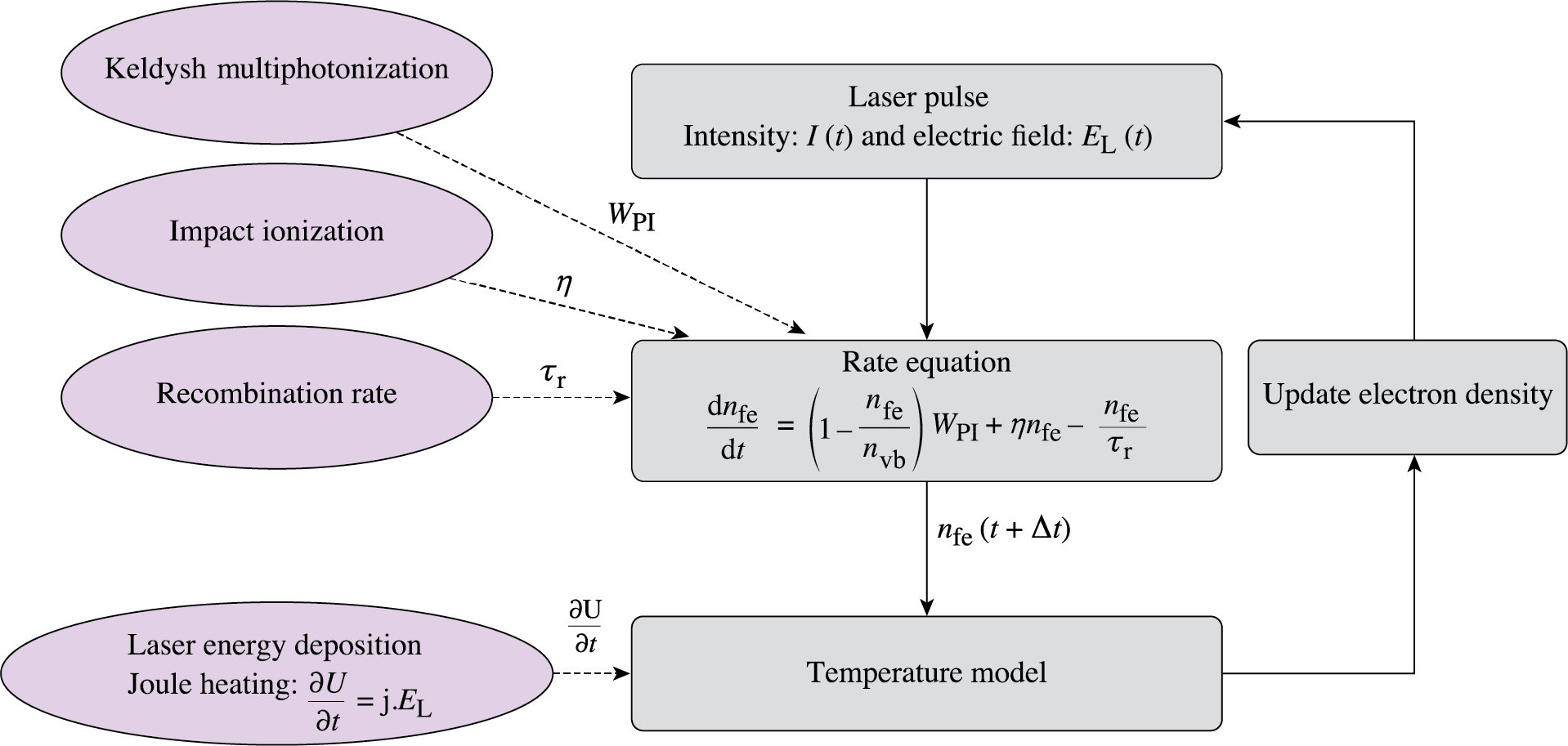}\quad
\caption{A flowchart detailing the different physical processes and the implementation sequence inside \LI.}
\label{flowchart}
\end{figure*}

The energy absorption mechanism changes based on the state of the matter. 
In the initial stages when the target is not ionized, the microphysics model dominates the energy absorption mechanism for the whole target. 
The energy deposited by the laser beam is absorbed by the free electrons through joule heating~\cite{Duchateau_Joule_Heating,Stuart_PRL}.
The incident energy density $U$ is calculated through a ray tracing algorithm that accounts for the angle of incidence of the laser beam, their intensity and geometry of the multiple beams with respect to the target.
As mentioned earlier, the electronic motion in the presence of an electric field is described by the Drude model. 
Therefore, the electrical conductivity of the medium from the Drude model is~\cite{Ginzburg}
\beq
\sigma=\frac{\sigma_0}{1-i\omega \nu^{-1}}, \quad \sigma_0=\frac{e^2n_{\te{fe}}}{m_{\te{e}}\nu}.
\eeq
This determines the free-electron current density in the medium, i.e., $\vec{j}=\sigma E_{\te{L}}$.
The laser energy density deposited per unit time heats up the free electrons in the medium through joule heating.
Specifically, the laser energy deposition is described by:
\begin{align}
\frac{\te{d}U}{\te{d}t}&=\text{Re}(\vec{j}.\vec{E_\te{L}})\nn\\
&=\frac{e^2 n_{\te{fe}}\nu}{m_{\te{e}}(\omega^2+\nu^2)}E_{\te{L}}^2.
\end{align}
Note that the energy deposition is proportional to the electron density. 
Thus, more energy is deposited in regions with higher electron density.
Eventually, the electron temperature is raised due to joule heating as the laser energy is deposited over time. 
Once the free electron density reaches the critical density for a given frequency of the incident light, the plasma is no longer transparent.
At this stage, the target is ionized in the coronal regions. 
The inverse bremsstrahlung absorption present in \LI\,is used for the coronal plasma behind the critical surface~\cite{Dawson}.
\Fig{flowchart} shows a flowchart outlining the sequence in which the rate equation and the laser-energy deposition scheme is implemented into the 1-D radiation hydrodynamic code \LI~\cite{Delettrez}.
The energy deposition mechanism is coupled with the two temperature model of \LI\,which is used to determine to the temperature profiles for the electrons and the ions. 
At each time step, the temperature profiles are updated based on the current electron and ion density which in turn changes based on the laser energy deposition.

\section{Results}\label{results}

\begin{figure}[htp]
\centering
\includegraphics[width=0.7\linewidth]{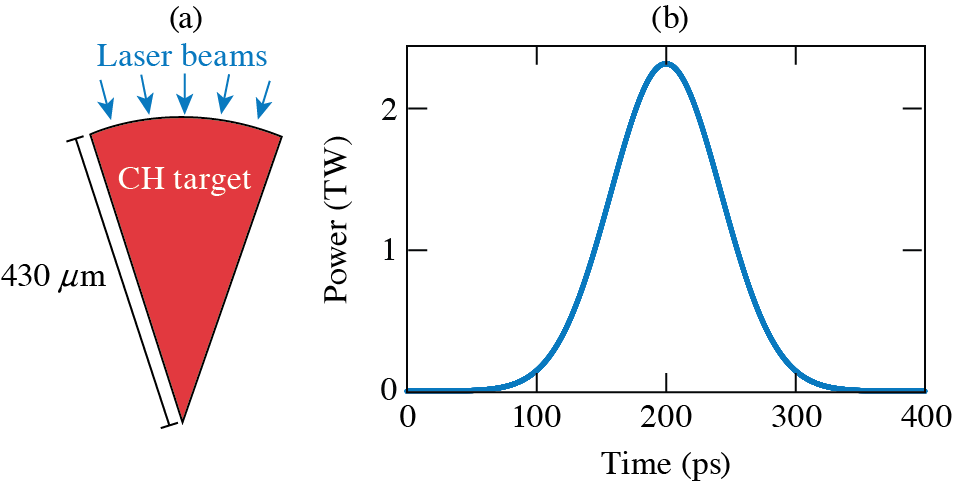}
\caption{(a) A solid plastic sphere of 430-$\mu$m radius is irradiated with (b) a laser pulse of 250-J energy.} 
\label{fig4}
\end{figure}

After we implemented this microphysics model into \LI, we examined how it affects hydrodynamic simulations in ICF.
Two such studies are given in this section.
For both the simulations, a time step of 0.1 fs was chosen such that it is smaller than the time scale at which any of the other physical processes can impact the electron density in the system.
The fixed and small time-step of 0.1 fs, chosen for accurate calculations of micro-processes for initial plasma formation, also guarantees that the Courant condition for hydro-motion is satisfied (with $C < 0.03$).”

\subsection{Plastic sphere}

The effect of irradiating a solid plastic sphere of 430-$\mu$m radius with a picket pulse (shown in \fig{fig4}) is discussed here.
The solid CH sphere is initially transparent to the UV light before the critical surface formation occurs around 81 ps according to simulation. 
The transition time to change from solid to plasma state is in the same order of magnitude as seen experimentally for other dielectric materials.~\cite{Stuart_PRB,Gamaly_1}
The microphysics model dominates the plasma profile for the entire sphere until the critical surface formation. 
Beyond that, the microphysics model controls the plasma profile ahead of the shock front. 

\begin{figure*}[!htpb]
\centering
\includegraphics[width=\linewidth]{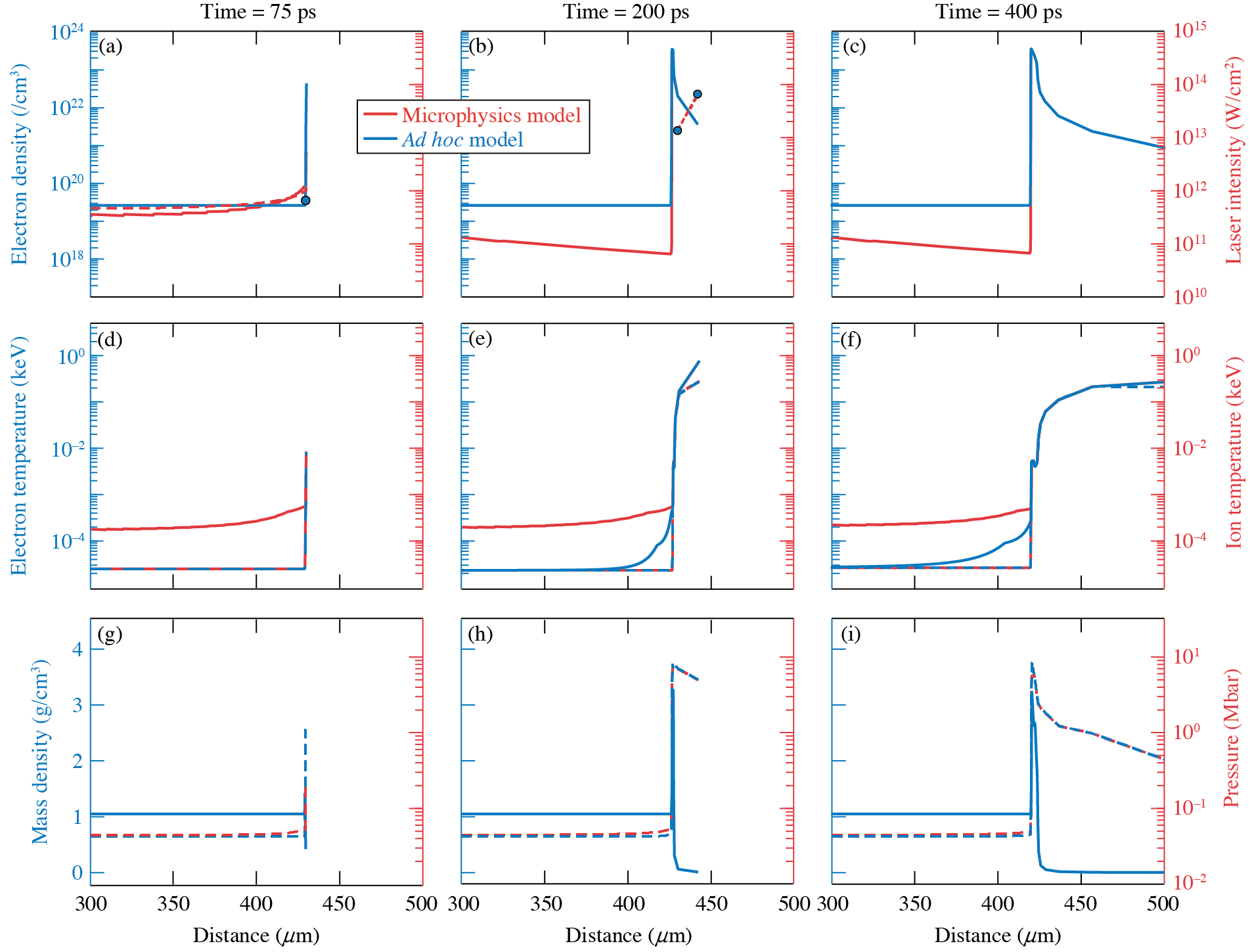}
\caption{The plasma profiles from the microphysics model and the \textit{ad hoc} model are plotted in red and blue respectively. The laser light is incident on the target from the right. The top row shows the laser intensity deposition profiles (in dashed red for microphysics model and solid blue circles for \textit{ad hoc} model) and the corresponding free electron density (in solid colors). The critical surface formation occurs when the free electron density rises to $9 \times 10^{21}$cm$^{-3}$ for UV light. The middle row shows the rise in the electron temperature (in solid) predicted by the microphysics model and the ion temperatures (in dashed). The mass density profile (in solid) and the difference in the pressure profiles (in dashed) is evident in the lowermost row.}
\label{fig5}
\end{figure*}

\Fig{fig5} shows the plasma profiles in the radially outward direction, 300 $\mu$m from the center of the sphere. 
The plasma profiles are plotted at 75 ps, i.e., before the critical surface formation, at the peak of the picket pulse (200 ps), and at 400 ps, which is the end of the picket pulse.
Earlier experiments have shown that CH is initially transparent to UV light~\cite{Edgell}.
The microphysics model predicts a nonzero laser-intensity profile inside the solid CH sphere that is consistent with the experimental observations.
Correspondingly, this leads to a rise in the free electron density due to photoionization, and electron temperature rises inside the target before the shock wave arrives.
Due to the rise in the electron temperature, the pressure inside the target also rises.
At 200 ps, the energy deposition occurs behind the critical surface around 430 $\mu$m.
No laser intensity is observed at the end of the laser pulse at 400 ps.
On the other hand, in the \textit{ad hoc} model of \LI, the energy deposition occurs in the last radial zone as marked by filled blue circles.
The rise in electron temperature and pressure is not observed in the \textit{ad hoc} model since the energy deposition is restricted to the surface of the target.
There is no observed difference in the ion temperature and mass-density profiles between the two models.
To summarize, the microphysics model predicts a rise in the electron temperature and pressure before the shock wave travels through the target due to the shinethrough mechanism of the laser light inside the CH.

\begin{figure}[!htpb]
\centering
\includegraphics[width=0.7\linewidth]{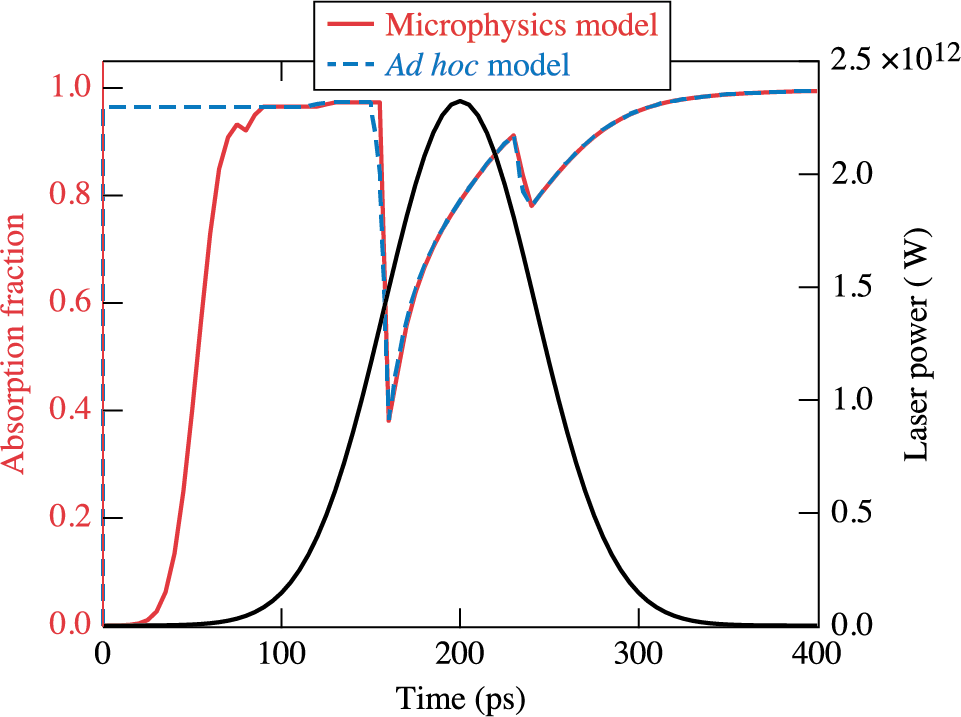}
\caption{The fraction of the incident laser energy absorbed over time. The absorption fraction between the microphysics model and the \textit{ad hoc} model is different initially since the energy deposition is initially restricted to the surface of the target for the \textit{ad hoc} model. Beyond the critical surface formation, the absorption profiles are the same since the plasma profile in the ablation region is controlled by the \textit{ad hoc} model.}
\label{fig6}
\end{figure}

\Fig{fig6} shows the fraction of incident energy absorbed over time. 
For the \textit{ad hoc} model, the incident energy is absorbed almost completely on the surface of the target leading to an absorption fraction of almost 1. 
In the microphysics model however, a small fraction of the incident energy is absorbed initially.
As the electron density grows, the energy absorbed increases until the critical surface formation, since the joule heating mechanism is proportional to the electron density.
This shine-through process occurs only before the establishment of critical density at the target surface. 
The early time shine-through creates free electrons in the target and heats up the target.
Eventually, the \textit{ad hoc} model and microphysics model have similar absorption fractions since the normal inverse bremsstrahlung absorption dominates the plasma profile behind the shock front or the ablation region after the critical surface formation.
For a real ICF target, the shinethrough mechanism can cause the pre-expansion of the ablator back-surface into the fuel and thereby affect the hydro-evolution in the later phase of laser-target interactions. 
Shock velocities in the target might be modulated due to this pre-expansion of the target-back surface.

\subsection{Planar plastic foil}

\begin{figure}[!htpb]
\centering
\includegraphics[width=0.7\linewidth]{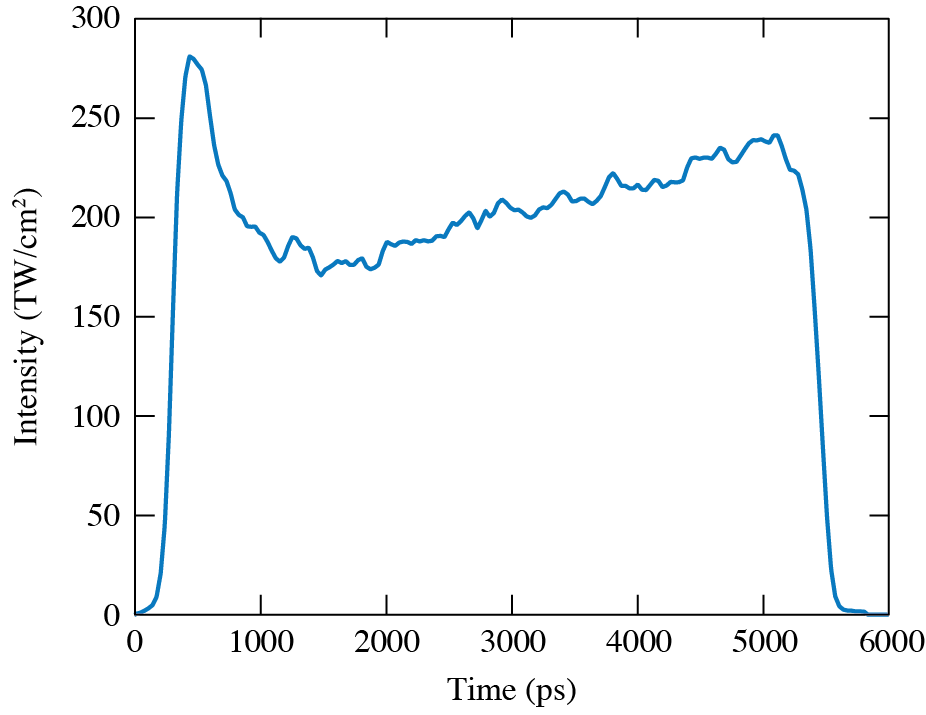}
\caption{Pulse shape used to irradiate a planar plastic foil of 37-$\mu$m thickness.}
\label{fig7}
\end{figure}

A recent experiment to study the material release from the inner shell of a target was conducted on the OMEGA EP laser~\cite{Haberberger}.
The experiment measured that the low-density plasma (at 10$^{20}$ cm$^{-3}$) released from the rear side of a laser-driven shell travelled a distance of $\sim$190 $\mu$m over 3 ns with a length scale longer than predictions from existing radiation-hydrodynamic codes.
To achieve good agreement between the radiation-hydrodynamic codes and the experiments, a relaxation of the back surface of the ablator that is consistent with experimental measurements had to be introduced in an arbitrary manner.
By incorporating the microphysics model however, the length scale for the material release was found to be larger than the \textit{ad hoc} model of \LI. 
The shinethrough mechanism provided by the microphysics model of initial plasma formation can possibly give rise to the needed pre-release of back surface of the CH foil without any arbitrary relaxation.
For this simulation, a planar foil of 37-$\mu$m thickness was irradiated with the square pulse shown in \fig{fig7}.

\begin{figure*}[!htpb]
\centering
\includegraphics[width=\linewidth]{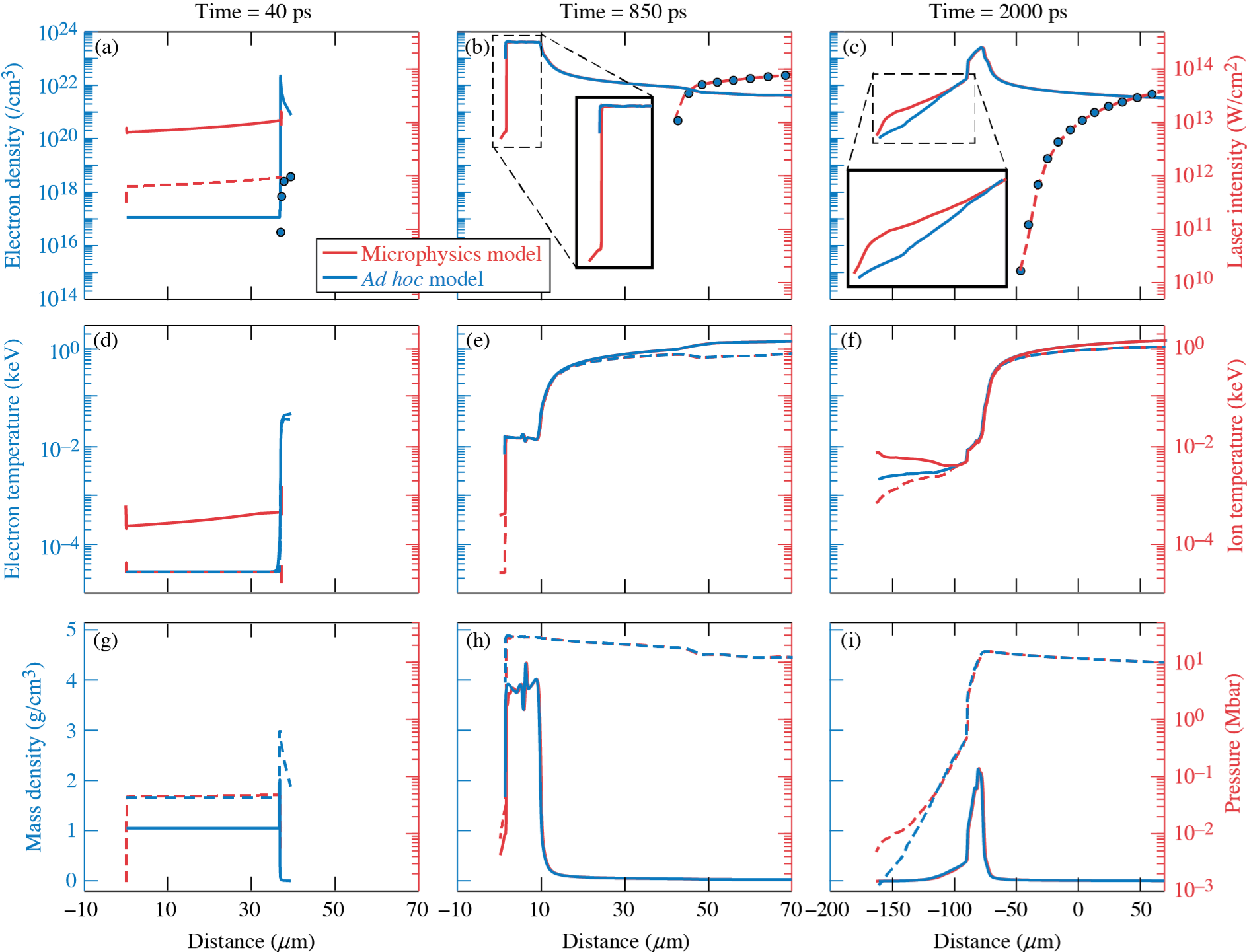}
\caption{The plasma profiles from the microphysics model and the \textit{ad hoc} model are plotted in red and blue respectively. The laser light is incident on the target from the right. The top row shows the laser intensity deposition profiles (in dashed red for microphysics model and filled blue circles for \textit{ad hoc} model) and the corresponding free electron density (in solid colors). The middle row shows the rise in the electron temperature (in solid) predicted by the microphysics model and the ion temperatures (in dashed). The mass density profile (in solid) and the difference in the pressure profiles (in dashed) is evident in the lowermost row. Note that the plastic is released in the rear end of the foil before the shock wave arrives due to the increase in the electron densities from the laser energy deposition. This leads to the material release over a larger length scale at 2 ns in the rear end of the foil as predicted by the microphysics model.}
\label{fig8}
\end{figure*}

\Fig{fig8} shows the plasma profiles obtained from the simulation at three different time steps: at 40 ps before the critical surface formation, at 850 ps when the shock front reaches the rear surface of the foil, and at 2 ns when the shock wave breaks out of the plastic foil.
The time taken for the critical surface formation in this simulation was 54 ps. 
The profiles show that due to energy deposition early in time, the free electron density, electron temperature, and pressure rise inside the planar foil for the microphysics model.
For this reason, the plastic in the rear end of the foil is released before the shock wave reaches the rear end. 
The increase in the electron density in the microphysics model is sustained even after the shock breaks out of the planar foil. 
The location of the higher electron densities $\sim10^{21}$cm$^{-3}$ at 2 ns shows that the material is released further into the rear end of the plastic foil in the microphysics model than the \textit{ad hoc} model.
This observation is consistent with the experimental observations suggesting that laser shinethrough could be the mechanism resulting in the decompression of the rear surface. 

\begin{figure}[!htpb]
\centering
\includegraphics[width=0.7\linewidth]{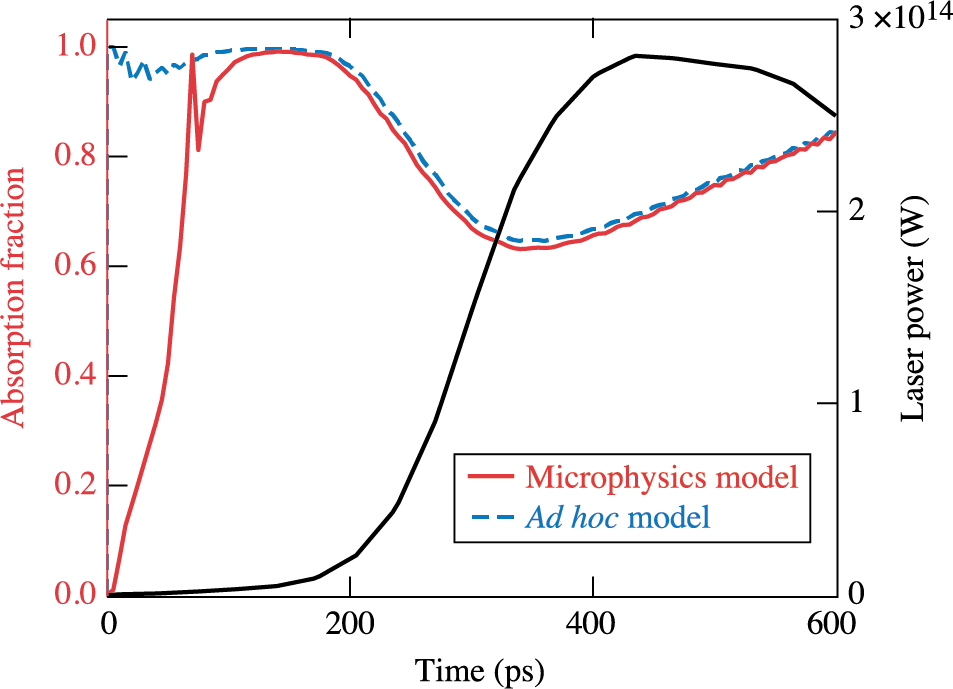}
\caption{The temporal distribution of the fraction of laser energy absorbed by the microphysics model shows lower absorption than the \textit{ad hoc} model initially until the critical surface formation. Over time, that the absorption profile has a similar trend as the distance between the critical surface from the two methods decreases and they overlap.}
\label{fig9}
\end{figure}

\Fig{fig9} shows that the absorption fractions are different initially for the microphysics model and the \textit{ad hoc} model and they become similar after the critical surface formation.
The shinethrough mechanism occurs until the establishment of critical density at the target surface. 
During this process, free electrons are created in the target that heats it up. 
This leads to the pre-expansion of the target back-surface and thereby affects the hydro-evolution in the later phase of laser-target interactions.
Over time, the distance between the location of the critical surfaces predicted by the two models decrease and they then overlap, leading to very similar absorption fractions.

\section{Conclusions}\label{summary}

A microphysics model to study the initial plasma formation has been implemented into the hydrodynamic code \LI\,for spherical and planar geometries. 
The results from the microphysics model demonstrate two new phenomena that had not been seen in numerical simulations earlier: (a) the shinethrough mechanism in plastic ablators, and (b) expansion of the rear end of a plastic target. 
During the shinethrough mechanism, free electrons are generated inside a target which heat up the target and lead to an increase in the electron temperature, and pressure in the transparent CH (ahead of the shock).
The pre-expansion of the target back-surface, on the other hand, affects the hydro-evolution in the later phase of laser-target interactions.
Eventually, a true test of the microphysics model would be a verification of the hypothesis that the rear surface of the plastic ablator expands into the DT ice inside a typical ICF target before the shock front reaches the ablator fuel interface.~\cite{Haberberger}
Before that, the next step is to implement the microphysics model into the 2-D hydrodynamic code \DR~\cite{Draco} for a more realistic investigation of the laser-imprint problem.
Perturbations to the ablation pressure as a function of angle due to the target response to laser imprint, will be modeled with \DR.
Efforts to study the consequences of the microphysics model for a cryogenic implosion are also underway as the material properties of DT gas and DT ice are being investigated.
It is necessary to know the band-gap, collisional frequency, and recombination rates for these materials to accurately implement the microphysics model.

\begin{acknowledgments}

We would like to thank V.~V. Karasiev and R. Epstein for useful discussions on this paper. 

This material is based upon work supported by the Department of Energy National Nuclear Security Administration under Award Number DE-NA0003856, the University of Rochester, and the New York State Energy Research and Development Authority. 

This report was prepared as an account of work sponsored by an agency of the U.S. Government. Neither the U.S. Government nor any agency thereof, nor any of their employees, makes any warranty, express or implied, or assumes any legal liability or responsibility for the accuracy, completeness, or usefulness of any information, apparatus, product, or process disclosed, or represents that its use would not infringe privately owned rights. Reference herein to any specific commercial product, process, or service by trade name, trademark, manufacturer, or otherwise does not necessarily constitute or imply its endorsement, recommendation, or favoring by the U.S. Government or any agency thereof. The views and opinions of authors expressed herein do not necessarily state or reflect those of the U.S. Government or any agency thereof.

\end{acknowledgments}

\appendix
\section{Keldysh formula for photoionization}\label{appen}

The Keldysh formula for the multiphoton ionization rate is~\cite{Keldysh}:
\begin{widetext}
\begin{eq}
W_{\te{PI}}&=\frac{2\omega}{9\pi}\left(\frac{\sqrt{1+\gamma^2}}{\gamma}\frac{m\omega}{\hbar}\right)^{\frac{3}{2}}Q\left(\gamma,\frac{\tilde{\Delta}}{\hbar\omega}\right)\exp\left\{-\pi\langle \frac{\tilde{\Delta}}{\hbar\omega}+1\rangle\right.\\
&\left.\times\left[K\left(\frac{\gamma}{\sqrt{\gamma^2+1}}\right)-E\left(\frac{\gamma}{\sqrt{\gamma^2+1}}\right)\right]/E\left(\frac{1}{\sqrt{\gamma^2+1}}\right)
\right\}
\end{eq}
where 
\begin{eq}
\tilde{\Delta}&=\frac{2}{\pi}\Delta\frac{\sqrt{1+\gamma^2}}{\gamma}E\left(\frac{1}{\sqrt{\gamma^2+1}}\right),\quad \gamma=\omega\frac{\sqrt{m_\te{e}\Delta}}{e E_\te{L}},\\
Q(\gamma,x)&=\left[\pi/2K\left(\frac{1}{\sqrt{1+\gamma^2}}\right)\right]^{\frac{1}{2}}\times \displaystyle{\sum_{n=0}^\infty}\exp\left\{-\pi\left[K\left(\frac{\gamma}{\sqrt{1+\gamma^2}}\right)-E\left(\frac{\gamma}{\sqrt{1+\gamma^2}}\right)\right] n/E\left(\frac{1}{\sqrt{1+\gamma^2}}\right)\right\}\\
&\times\Phi\left\{\left[\pi^2(2\langle x+1\rangle-2x+n)/2K\left(\frac{1}{\sqrt{1+\gamma^2}}\right)\times E\left(\frac{1}{\sqrt{1+\gamma^2}}\right)\right]^{\frac{1}{2}}\right\}.
\end{eq}
\end{widetext}
Here $\langle x\rangle$ denotes the integer part of $x$; $K(x)$ and $E(x)$ are the complete elliptic integrals of the first and second kind. 
In this expression, $Q(\gamma,x)$ captures the discrete nature of the number of absorbed photons associated with the spectrum or band-gap of CH.
The $W_{\te{PI}}$ term accounts for both the multiphoton absorption process for $\gamma\gg1$ and tunneling for $\gamma \ll 1$~\cite{Keldysh,Zheltikov_review}.

\bibliography{references}

\end{document}